\documentclass[english,twocolumn,aps,prl, superscriptaddress,showpacs, amsmath]{revtex4-1}

\usepackage{color}

\usepackage{natbib}
\usepackage[T1]{fontenc}
\usepackage[latin9]{inputenc}
\usepackage{units}

\usepackage{amssymb}
\usepackage{braket}
\usepackage{graphicx}
\usepackage{subcaption}
\usepackage{esint}
\usepackage{bm}
\usepackage{ulem}
\usepackage[colorlinks=true, citecolor=blue, urlcolor=blue]{hyperref}
\setcitestyle{journalcolor= blue}
\usepackage{lineno}
\usepackage{afterpage}

\begin{document}

\title{Imaging orbital Rashba induced charge transport anisotropy}
\author{Eylon Persky}
\affiliation{Department of Physics, Bar-Ilan University, Ramat-Gan 5290002, Israel}
\affiliation{Institute of Nanotechnology and Advanced Materials, Bar-Ilan University, Ramat-Gan 5290002, Israel}
\author{Xi Wang}
\affiliation{Department of Physics, Bar-Ilan University, Ramat-Gan 5290002, Israel}
\affiliation{Institute of Nanotechnology and Advanced Materials, Bar-Ilan University, Ramat-Gan 5290002, Israel}
\author{Giacomo Sala}
\affiliation{Department of Quantum Matter Physics, University of Geneva,
24 Quai Ernest Ansermet, CH-1211 Geneva, Switzerland}
\author{Thierry C. van Thiel}
\affiliation{Kavli Institute of Nanoscience, Delft University of Technology, Lorentzweg 1, 2628CJ Delft, Netherlands}
\author{Edouard Lesne}
\affiliation{Kavli Institute of Nanoscience, Delft University of Technology, Lorentzweg 1, 2628CJ Delft, Netherlands}
\affiliation{Max Planck Institute for Chemical Physics of Solids, 01187 Dresden, Germany}
\author{Alexander Lau}
\affiliation{International Research Centre MagTop, Institute of Physics, Polish Academy of Sciences, Aleja Lotnik\'{o}w 32/46, PL-02668 Warsaw, Poland}
\author{Mario Cuoco}
\affiliation{Consiglio Nazionale delle Ricerche, CNR-SPIN, Italy}
\author{Marc Gabay}
\affiliation{Laboratoire de Physique des Solides, Universite Paris-Saclay, CNRS UMR 8502, Orsay, France}
\author{Carmine Ortix}
\email{cortix@unisa.it}
\affiliation{Dipartimento di Fisica "E. R. Caianiello", Universit\'{a} di Salerno, IT-84084 Fisciano, Italy}
\author{Andrea D. Caviglia}
\email{andrea.caviglia@unige.ch}
\affiliation{Department of Quantum Matter Physics, University of Geneva,
24 Quai Ernest Ansermet, CH-1211 Geneva, Switzerland}
\author{Beena Kalisky}
\email{beena@biu.ac.il}
\affiliation{Department of Physics, Bar-Ilan University, Ramat-Gan 5290002, Israel}
\affiliation{Institute of Nanotechnology and Advanced Materials, Bar-Ilan University, Ramat-Gan 5290002, Israel}

\begin{abstract}
Identifying orbital textures and their effects on the electronic properties of quantum materials is a critical element in developing orbitronic devices. However, orbital effects are often entangled with the spin degree of freedom, making it difficult to uniquely identify them in charge transport phenomena. Here, we present a combination of scanning superconducting quantum interference device (SQUID) current imaging, global transport measurements, and theoretical analysis, that reveals a direct contribution of orbital textures to the linear charge transport of 2D systems. Specifically, we show that in the LaAlO$_3$/SrTiO$_3$ interface, which lacks both rotation and inversion symmetries, an anisotropic orbital Rashba coupling leads to conductivity anisotropy in zero magnetic field. We experimentally demonstrate this result by locally measuring the conductivity anisotropy, and correlating its appearance to the non-linear Hall effect, showing that the two phenomena have a common origin. Our results lay the foundations for an all--electrical probing of orbital currents in two-dimensional systems.
\end{abstract}

\maketitle
\section{Introduction}
The multi-orbital nature of oxide interfaces \cite{Pai2018, Hwang2012,Ahn2021}, combined with their low crystalline symmetry, enables a plethora of quantum phenomena, such as quantum geometry induced non-linear transport \cite{lesne2022designing,Mercaldo_2023,Sala2024} and unconventional superconductivity \cite{Reyren2007,Caviglia2008}. 
The gate tunability of these properties is particularly promising as a platform to engineer novel devices or investigate transitions between different electronic phases. One example is the orbital Hall effect (OHE) \cite{Bernevig_2005} and its surface counterpart, the orbital Rashba-Edelstein effect (OREE), where a charge current generates a transverse flow of orbital angular momentum. These effects are analogous to the spin Hall effect and the spin Rashba-Edelstein effect, respectively. Through this effect, the orbital degree of freedom can be used to generate and detect information \cite{Go_2021A}. Both the OREE and the OHE can be directly related to the existence of momentum-space orbital textures \cite{Go_2018}. In non-centrosymmetric low-dimensional materials these textures arise from
a linear coupling, known as the orbital Rashba coupling \cite{Park_2011, Sunko_2017}, between the crystalline momentum ${\bf k}$ and the orbital angular momentum ${\bf L}$. The coupling strength can be controlled using electrostatic fields, setting the stage for a controlled generation of orbital currents. This provides a strong motivation to detect orbital textures and identify their sources in materials with low crystalline symmetry.
One approach is to analyze how such couplings affect the charge transport properties of the material. For example, Rashba SOC gives rise to weak anti-localization, with signatures such as positive magnetoresistance \cite{Hikani1980}and (non) linear transport due to band geometric properties \cite{Moore2010,Sodemann_2015,lesne2022designing,Sala2024}. In contrast, the signatures of orbital effects are often difficult to detect, because they are weak, or because multiple microscopic mechanisms can explain the observed experimental results. In these cases, new tools, or a combination of several probes, are required to pinpoint the cause of transport anomalies.

A notable example is the transport anisotropy in LaAlO$_3$/SrTiO$_3$. At low temperatures, the tetragonal distortion of SrTiO$_3$ breaks the rotational symmetry, allowing for anisotropy in the DC, zero-field charge transport. Indeed, experiments generally measure such anisotropy \cite{BenShalom_2009,Annadi_2013,Frenkel2016,Goble2017,Krantz2021,Prasad2021}, but its source has been disputed. Several intrinsic mechanisms can generate anisotropy, including Fermi surface warping and the orbital Rashba coupling. The latter option, which further requires inversion symmetry breaking, is particularly interesting because it gives rise to transverse orbital currents, which are useful for device applications \cite{Go_2021,BurgosAtencia2024}. In the superconducting state, such orbital mixing could also lead to new pairing channels \cite{mercaldo2023aqt}. However, the large variability in reported results, as well as several scanning probe experiments \cite{Kalisky2013,Honig2013,Ma_2016,Frenkel2016,Noad2016,Wissberg2017,Goble2017,Persky2021}, suggested that the origin could be extrinsic. Namely, tetragonal domains of different orientations can have different properties \cite{Wissberg2017,Noad2016}, or the walls between the structural domains could host highly-conducting quasi-1D channels \cite{Frenkel2017,Cheng2018,Pai2018a} that manifest as anisotropy in the global transport measurement. Determining the origin of anisotropy is crucial to interpreting quantum transport measurements in SrTiO$_3$-based interfaces, and for harnessing their properties for device applications.

In this work, we use a combination of global transport measurements, local current imaging and theoretical analysis to associate the origin of transport anisotropy in (111) oriented LaAlO$_3$/SrTiO$_3$ interface with the orbital Rashba coupling. Firstly, we locally probe the current flow at the interface and show direct and unambiguous evidence that the anisotropy is an intrinsic property of the interface. Secondly, we measure the non-linear Hall effect, whose presence is a direct indicator of inversion symmetry breaking. We find a correlation between the temperature dependence of the two effects, which suggests that they have a common origin. Finally, we interpret these results using a $\mathbf{k}\cdot\mathbf{p}$ model of the band structure, which allows us to identify the anisotropic orbital Rashba coupling - a linear-in-momentum, anisotropic mixing of the $t_{2g}$ orbitals - as the leading term responsible for both effects. Our results provide a new platform for exploring orbital effects and controlling the orbital degree of freedom in various systems.
\section{Results}
\subsection{Intrinsic anisotropy in (111) LaAlO$_3$/SrTiO$_3$}

To distinguish between intrinsic and extrinsic origins of transport anisotropy, we used a local probe, a scanning superconducting quantum interference device (SQUID), to directly image the current flow at the interface. Scanning SQUID was previously used to study (001)-based interfaces, revealing that current density is spatially inhomogeneous, and that the inhomogeneity is correlated with the tetragonal domain structure of the SrTiO$_3$ substrate \cite{Kalisky2013}. While these observations might point to an extrinsic source - the network of domains or domain walls - this determination cannot be made based on existing data. Specifically, it was difficult to distinguish between enhanced current density on entire domains versus domain boundaries, and to relate these to the components of the conductivity tensor. Here we study the (111) oriented interface, in which we found that domains tend to be wider, and the tiling rules of the domains enable us to determine their orientations unambiguously.

\begin{figure*}[ht!]
\centering
\includegraphics[scale=1]{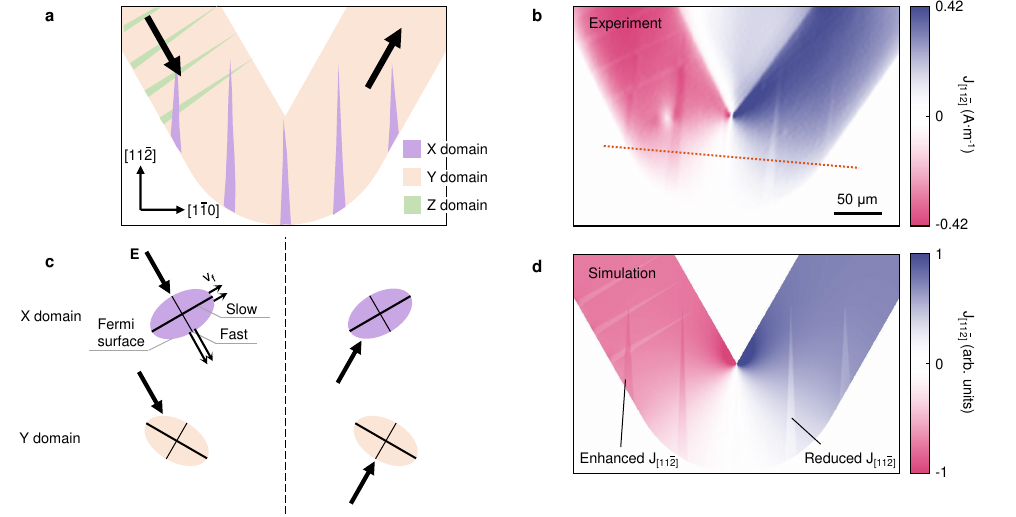}
\caption{Modulation reversal upon reflection of the current flow. (a) Schematic of a triangle-shaped device with an X-Y domain pattern. The current density is reflected around the $[11\bar{2}]$ direction upon flowing across the device. (b) A scanning SQUID image of the current flow in a $(111)$-oriented device. The current is modulated over the domain pattern: on the left side of the device, current density on X domains is larger than on Y-domains. This trend is reversed on the right side of the device, coinciding with a reflection of the overall current density. The modulation strength is about 8\%. (c) Schematic of the Fermi surface at the X and Y domains, showing the direction of the electric field on the two sides of the device. On the left, the electric field is aligned with the fast direction on X domains, resulting in increased current density. This is reversed on the right side of the device.
(d) Simulated current density map of the device, where an anisotropic conductivity tensor was introduced in order to account for the reversed modulations.}
\label{reversed modulation}
\end{figure*}

Figure \ref{reversed modulation} shows results from a triangle patterned (111)-oriented LaAlO$_3$/SrTiO$_3$ device (Methods). We observed current density modulations (Figure \ref{reversed modulation}b), across the device. The modulations changed after a thermal cycle above 105 K, confirming that they originate from the STO tetragonal domains (Fig. \ref{T cycle}). Based on the tiling rules, we determined the orientation of the domains to be along the [100] and [010] axes.
The modulations were over 10 $\mu$m wide strips which terminate at needle-like shapes. 
The modulations reveal a surprising dependence on the device geometry: On the left side of the triangle, the current density along $[11\bar{2}]$ is larger on X domains than on Y domains but lower on the right side. The change coincides with a reflection of $\mathbf{J}$ with respect to the $[11\bar{2}]$ crystallographic direction. This relationship between the current direction and the modulation has not been previously reported in the (001) interface. We thus associate the modulations with a change in the properties of the X and Y domains, rather than a domain-boundary effect. Figs. \ref{example 1}, \ref{example 2} and \ref{example 3} show data from 3 additional devices with various geometries and domain patterns.

To understand the difference between the (001) and (111) interfaces, we note that at (001) interfaces, (001)-oriented domains (Z-domains) have a ${\mathcal C}_{4v}$ symmetry, while (100) and (010) domains (X,Y-domains) have ${\mathcal C}_{2v}$ symmetry. 
Therefore, in (001) interfaces, the conductivity of Z domains is intrinsically isotropic, and is described by a single scalar. For X and Y domains, the conductivity tensor has a symmetry-allowed anisotropy, but it is diagonal with respect to the (100) and (010) directions. The situation changes in (111) oriented interfaces. All domains have the same crystalline symmetry group, ${\mathcal C}_{s}$, which admits an anisotropic conductivity tensor with principal directions that do not coincide with the crystallographic directions.
Denoting the current density as the 2D vector
\begin{equation}
    \mathbf{J} = \begin{pmatrix}
    J_{[1\bar{1}0]}\\
    J_{[11\bar{2}]}
    \end{pmatrix},
\end{equation}
The conductivity tensor for Z-domains is anisotropic but diagonal,
\begin{equation}
    \sigma^{Z} = \begin{pmatrix}
    \sigma_{xx} & 0\\
    0 & \sigma_{yy}
    \end{pmatrix},
\end{equation}
while the X and Y domain conductivities are related by a $120^{\circ}$ rotation, 
\begin{equation}
        \sigma^{X,Y} = \frac{1}{4} \begin{pmatrix}
    \sigma_{xx} + 3\sigma_{yy} & \pm\sqrt{3} (\sigma_{xx} - \sigma_{yy})\\
    \pm\sqrt{3} (\sigma_{xx} - \sigma_{yy}) & \sigma_{yy} + 3\sigma_{xx}
    \end{pmatrix}.
\end{equation}

Note that the off-diagonal elements are non-zero (in zero magnetic field) only if the conductivity is anisotropic ($\sigma_{xx} \neq \sigma_{yy}$).

The off-diagonal components of the conductivity are at the origin of the reversed modulations observed in the data. In Fig. \ref{reversed modulation}c, we schematically show an anisotropic Fermi surface with corresponding direction-dependent Fermi velocities that yield a fast and a slow direction. The current density $\mathbf{J}$ is increased when the electric field is aligned with the fast direction of the Fermi surface, which causes the modulations. On the left side of the device, this alignment is greater in x domains. As the current flow is reflected about the $[11\bar{2}]$ direction, due to the triangle shape of the device, $\mathbf{E}$ is aligned with the fast direction in the Y domains, thus reversing the current density modulation. Fig. \ref{reversed modulation}d shows finite element simulations (Methods) of the current density in such geometry, which confirm that the modulation trends are reversed. Furthermore, domains with different (but isotropic) conductivity would lead to modulations that do not reverse sign upon reflection of the device (Methods). Thus, observing this change of sign is a direct measure of the intrinsic conductivity anisotropy of an individual domain.

\begin{figure}[ht]
\centering
\includegraphics{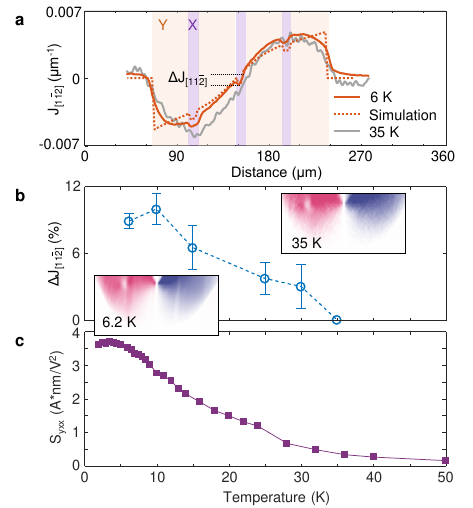}
\caption{Temperature and carrier density dependence of the anisotropy. (a) Line-cut along the dashed line in Fig. \ref{reversed modulation}b, along with a line cut of the simulated current densities. The modulation is 10\%. The gray line is a line-cut taken at the same location but at 35 K. The current densities are normalized by the applied current. (b) Temperature dependence of the current density modulations shown in panel a. The modulation of current density decreases with temperature and disappears near 35 K. Insets: current density maps at 6 K and 35 K. (c) Temperature dependence of the non-linear conductivity, measured at zero magnetic field, showing similar onset temperature to the anisotropy.
}
\label{T-dep}
\end{figure}

\subsection{Correlation between anisotropy and inversion symmetry breaking}

To investigate the origin of the anisotropy, we studied the temperature dependence of the modulations. Fig. \ref{T-dep}a,b shows how the current density modulations decreased upon increasing the temperature, until disappearing at ~35 K. This temperature dependence is consistent with previous observations on the (001) interface \cite{Kalisky2013}. It is important to note that even though the domains exist already at 105 K, and the crystalline symmetry allows anisotropy to exist below 105 K, it is not observed in the range $35 < T < 105$, suggesting there is another mechanism which generates, or enhances the anisotropy at lower temperatures.

There are several possible mechanisms. 
Scattering can smear the effect at elevated temperatures, explaining the gradual change in the signal. 
In SrTiO$_3$, the scattering could be suppressed at low temperatures due to the large dielectric constant. 
Alternatively, the appearance of modulations only below 35 K can suggest that the modulations are caused by a structural change occuring at this temperature. 
Bulk SrTiO$_3$ shows an increased dielectric permittivity below 40 K, signalling an incipient ferroelectric state \cite{Muller1979}. This quantum paraelectric phase significantly enhances the inversion symmetry breaking that is already present at the interface \cite{Zhang2023}. 

To understand the source of the anisotropy, we further studied the symmetry of the system by measuring the nonlinear Hall effect. Observing nonlinear transport is a direct consequence of inversion symmetry breaking. Furthermore, previous work on oxide interfaces, which studied the magnetic field dependence of the nonlinear Hall \cite{lesne2022designing}, associated it with a Berry curvature dipole, which indicates the presence of an orbital Rashba coupling. We measured the temperature dependence of the nonlinear Hall conductivity on a four-point device on the same sample, and found that it onsets at 40 K (Fig. \ref{T-dep}c). This temperature is in line with previous work showing the same characteristic temperature \cite{lesne2022designing}. Remarkably, both the linear conductivity anisotropy and the nonlinear Hall conductivity onset at the same temperature, even though the corresponding crystalline symmetries (rotation and inversion) are already broken at much higher temperatures. This correlation indicates that the two effects share a common origin.

\subsection{$\mathbf{k}\cdot\mathbf{p}$ model and sources of anisotropy}
We use a $\mathbf{k}\cdot\mathbf{p}$ model to analyze the symmetry allowed terms and their contribution to the transport anisotropy. Above 105 K, the system retains its trigonal symmetry, and belongs to the $\mathcal{C}_{3v}$ symmetry group. At zero magnetic field, the spin degree of freedom does not contribute to the linear conductivity \cite{Brosco_2016}. We therefore consider a  spinless Hamiltonian, which contains the following symmetry-allowed terms: (1) Splitting between the $a_{1g}$ singlet and the $e_g^{\pi}$ doublet due to the trigonal field, (2) $|\mathbf{k}|^2$ terms that account for the trigonal fermi surface warping. They give rise to an orbital-dependent effective mass, as observed in ARPES experiments \cite{Rodel_2014} and (3) $k$-linear terms that are allowed by the broken inversion symmetry. These terms, which are proportional to $\mathbf{k}\cdot\mathbf{L}$ where $\mathbf{L}$ are the angular momentum matrices, are the orbital Rashba coupling. They mix the orbital bands and give rise to transverse orbital currents and Berry curvature dipoles.

Below 105 K, the rotational symmetry is broken, reducing the symmetry group to $\mathcal{C}_s$ and enabling a plethora of new terms in the Hamiltonian. These terms can be casted into three categories (Methods):
(4) splitting of the $e_g^\pi$ bands, (5) an anisotropic and orbital dependent effective mass (terms proportional to $k_x^2$ or $k_y^2$) and \cite{Bednyakov2015,McKeown_2014,Syro_2012,King_2012} (6)
an anisotropic orbital Rashba coupling (linear in $k_x$ and $k_y$).

\begin{figure}[ht]
\centering
\includegraphics[scale=1]{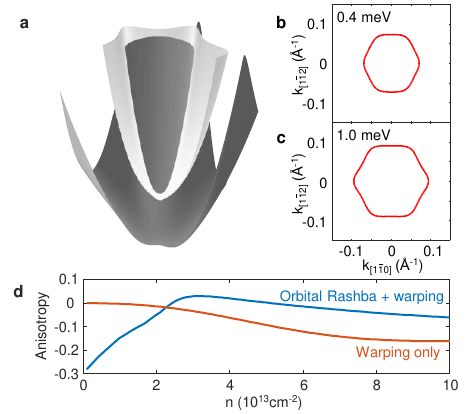}
\caption{The anisotropic band structures of (111) LaAlO$_3$/SrTiO$_3$, and carrier density dependence of the anisotropy. (a) The band structure calculated from an effective $\mathbf{k}\cdot \mathbf{p}$ model, showing effects of the orbital Rashba coupling (lateral splitting of the lowest band), Fermi surface warping (flower-like shape at higher energies) and the tetragonal splitting of the $e_g^{\pi}$ bands. (b,c) The Fermi surfaces 0.4 meV and 1 meV, showing the anisotropy at low energies, where the band structure is still dominated by the linear (Rashba) terms. (d) Calculated anisotropy as a function of the carrier density with and without including the orbital Rashba coupling.}
\label{band structure}
\end{figure}

The correlation between the non-linear Hall effect (broken inversion) and the anisotropic linear transport (broken rotation) suggests that the two effects are intertwined. Particularly, we expect the orbital Rashba terms to change strongly at 40 K, as indicated by the non-linear Hall. Figure \ref{band structure}a shows the band structure obtained by fitting to photoemission data from a (111) SrTiO$_3$ surface \cite{Rodel_2014}. Since the ARPES data fits well to an isotropic model, it provides upper bounds on the contributions from the $\mathcal{C}_s$ terms discussed above. To account for the tetragonal distortion, we only include a small splitting (1 meV) of the $e_g^{\pi}$ bands. Figure \ref{band structure}a-c shows the resulting band structure and the corresponding Fermi surfaces at carrier densities relevant to our experiment (Fig. \ref{Magnetotransport}). The band structure has several important features. Firstly, at low densities it is dominated by the orbital Rashba coupling, which gives rise to the two shifted parabolic features in the lowest bands. Note that at these low energies, the Fermi surface does not show significant tetragonal warping (Fig. \ref{band structure}b), but is strongly anisotropic. At higher energies the warping becomes significant. Notably, the anisotropy in the higher $e_g^{\pi}$ bands is opposite to that of the $a_{1g}$ band (Fig. \ref{band structure}a). Correspondingly, the transport anisotropy computed using the Boltzmann formalism (see Methods) is non-monotonic and is stronger at low densities (Fig. \ref{band structure}d). This matches the behavior observed in experiment over the smaller (experimentally attainable) range of densities (Fig. \ref{Gate dep images 2}.)

To assess the contribution of Fermi surface warping to the transport anisotropy, we repeated the calculation without including the orbital Rashba effect. The resulting anisotropy was zero at low carrier densities and varied monotonically and quadratically with increasing densities, reflecting the contributions from the quadratic terms. These results indicate that the tetragonal splitting of the $e_{g}^{\pi}$ bands on its own is not sufficient to explain the anisotropy observed in the experiment. While an anisotropic quadratic (mass) term is allowed, and can result in a non-monotonic dependence on the carrier density, the combination of the SQUID images, non-linear Hall effect, temperature and gate dependence indicate that the anisotropy is correlated with the broken inversion symmetry, consistent with the calculations we present, that indicate that the orbital Rashba coupling greatly enhances the anisotropy at low energies.


\section{Discussion}

In this work, we show that momentum space orbital textures can be probed by charge transport by lowering the crystalline symmetry. Specifically, the character of the orbital Rashba coupling in the absence of rotational symmetries results in a specific intra-atomic contribution to the orbital Hall effect that is absent in more symmetric crystalline environments. This provides a direct correspondence between charge transport anisotropy and the occurrence of orbital currents causing the OHE.
We developed an approach to explore the anisotropy by extracting it from local measurements of the current density, and demonstrated it on (111) LaAlO$_3$/SrTiO$_3$.

In LAO/STO, spin-orbit coupling plays an important role in determining the band structure and electronic behavior \cite{Caviglia2010,BenShalom2010,Joshua2012,Rout2017,Monteiro2019}. The zero-field intrinsic anisotropy uncovered here demonstrates our ability to separate the orbital and spin contributions to transport and identify new features in the band structure. Furthermore, the local measurements enabled us to study the interplay between the orbital currents and the crystalline tetragonal domain patterns. We identified them as the source of the current density modulations previously observed in LAO/STO \cite{Kalisky2013,Honig2013,Frenkel2016,Frenkel2017,Goble2017,Persky2021}. We showed that the modulations are caused by anisotropic conductivity, whose principle axes are rotated across different domains.

More generally, our results indicate that by choosing the appropriate crystalline symmetry, orbital currents can be identified by measuring transport anisotropy. This task can be accomplished via various local techniques \cite{Casola_2018,Ella_2019,Muralt_1986,Martin_2007,Waissman_2013} and also through measurements of the global transport anisotropy. Uniquely identifying orbital Hall currents, for example, has been a challenging task because their signatures are similar to spin Hall currents, and therefore limited to systems with low spin-orbit coupling \cite{Sala2023,Choi_2023}. Our approach offers a way to disentangle the spin and orbital contributions, thereby allowing the study of orbital phenomena in materials with strong SOC.

Finally, we note that the orbital effects we observed in LaAlO$_3$/SrTiO$_3$ are closely associated with the low-temperature transitions of SrTiO$_3$. However, a similar low-symmetry structure could be engineered at room temperature by breaking the rotation and inversion symmetries using epitaxial strain and ferroelectric layers. This engineering of orbital-induced anisotropy is relevant not only to complex oxide interfaces. It applies to other orbitronics material candidates, including light and heavy elemental metals and van der Waals heterostructures. The theoretical model and experimental techniques presented here provide an effective platform to characterize the orbital degrees of freedom in these systems.

\section{Materials and Methods}
\setcounter{equation}{0}    
\subsection{Sample growth and device fabrication}
To synthesize the LaAlO$_3$/SrTiO$_3$ heterostructure, a commercially available $(111)$ SrTiO$_3$ substrate with a Ti-rich surface was pasted onto a metal holder and heated to $740^{\circ}\mathrm{\,C}$, at a rate of $15^{\circ}\mathrm{\,C\, min^{-1}}$ in an O$_2$ pressure of $6\times 10^{-5} \mathrm{\,mbar}$. 
An 11 unit cell thick LaAlO$_3$ layer was subsequently grown through pulsed-laser deposition, using a $248 \mathrm{\,nm}$ KrF excimer laser at a $1 \mathrm{\, Hz}$ repetition rate and a fluence of $1 \mathrm{\,J\,cm^{-2}}$. The sample surface was monitored during the growth using in-situ reflection high-energy electron spectroscopy (RHEED). 

Device geometries were defined by RF-sputtering a HfO$_2$ hard mask in areas meant to remain insulating, prior to the growth of the LaAlO$_3$ layer. The hard mask was defined by electron-beam lithography, using a double-layer PMMA liftoff resist stack.

\subsection{Imaging currents with scanning SQUID microscopy} The sample was glued on a conductive surface and electrical contacts made by ultrasonic wire bonding. A back gate voltage was applied between the interface and the metallic sample holder. To locally measure the current density, an alternating current (34 $\mu$A RMS, frequency 1163 Hz for the triangle-shaped device shown in Fig \ref{reversed modulation}; 40 $\mu$A RMS, frequency 1009 Hz for the device shown in Fig. \ref{current inversion 2} and \ref{modulation}) was applied to the LaAlO$_3$/SrTiO$_3$ device. The sample was scanned over a planar SQUID sensor with a 0.8 $\mu$m diameter pick-up loop \cite{Huber_2008}. The sensor-sample distance was kept at 1 $\mu$m. The magnetic flux due to the current flow was recorded using a lock-in amplifier.

Fourier analysis was used to reconstrct the current density from the measured magnetic flux \cite{Roth1989,Nowack2013}. First, the out of plane component of the magnetic field, $B_z(x,y)$ was recovered from the data by deconvolution with the sensor's point spread function.

The magnetic field generated by $\mathbf{J}(\mathbf{r}) = J_x(\mathbf{J})\mathbf{\hat{x}}+J_y(\mathbf{J})\mathbf{\hat{y}}$ is given by the Biot-Savart law,
\begin{equation} \label{eq-BiotSavart}
    \mathbf{B} (\mathbf{r}) = \frac{\mu_0 d}{4\pi}\int \mathrm{d}x'\int \mathrm{d}y' \frac{\mathbf{J}(\mathbf{r'}) \times (\mathbf{r} - \mathbf{r'})}{|\mathbf{r} - \mathbf{r'}|^3},
\end{equation}
where $\mu_0$ is the vacuum permeability and $d$ is the thickness of the 2D layer. Equation \ref{eq-BiotSavart} is a convolution integral. Therefore, the two-dimensional Fourier transform of the out-of-plane component of the magnetic field,
\begin{equation}
    B_z(k_x,k_y,z) = \int \mathrm{d}x \int \mathrm{d}y B_z(x,y,z)e^{ik_xx+ik_yy},
\end{equation}
can be written as
\begin{equation} \label{eq-BS_FT}
    \tilde{B}_z (k_x,k_y,z) = -i\frac{\mu_0d}{2}e^{-|k|z} \left[ \frac{k_y}{|k|}\tilde{J}_x(k_x,k_y) - \frac{k_x}{|k|}\tilde{J}_y(k_x,k_y) \right],
\end{equation}
where $|k| = \left(k_x^2+k_y^2\right)^{1/2}$, and $\tilde{J}_x$ and $\tilde{J}_y$ are the Fourier transforms of $J_x$ and $J_y$, respectively. Equation \ref{eq-BS_FT} was used together with the continuity equation, $\mathbf{k}\cdot\tilde{\mathbf{J}} = 0$, to extract $J_x$ and $J_y$ from $B_z$. Fig. \ref{current inversion 2} shows a representative magnetic flux image recorded by the SQUID and the $x$ and $y$ components of the reconstructed current density. All the features seen in the reconstructed images are clearly visible in the raw data as well.

\subsection{Identification of domain orientations}
Domain orientations were determined by examining the tiling rules established at domain intersections. Fig. \ref{modulation}a shows the possible orientations of the tetragonal domain boundaries in STO, projected onto the $(111)$ surface \cite{Bednyakov2015}. Each pair of domain orientations can form a boundary oriented at one of two angles with respect to the $[\bar{1}10]$ direction. For example, Fig. \ref{modulation}b shows a representative current density map of a device oriented $60^{\circ}$ with respect to the $[1\bar{1}0]$ direction. The current modulations reveal three types of regions: a set of narrow ($10 \mathrm{\,\mu m}$) stripes of lower current density oriented along the device, the wider regions of higher current density between them, and a set of wide ($30 \mathrm{\,\mu m}$) stripes with lower current density, oriented $30^{\circ}$ with respect to $[1\bar{1}0]$ direction. The orientations of these regions and the intersections between them suggest that these regions coincide with X,Y,Z tetragonal domains, respectively.

\subsection{Current density simulations}
To simulate the current density resulting from an anisotropic conductivity tensor, the continuity equation
\begin{equation}
    \nabla \cdot \mathbf{J} = 0,
\end{equation}
and Ohm's law
\begin{equation}
    \mathbf{J} = \hat{\sigma}\mathbf{E},
\end{equation}
where $\hat{\sigma}$ is the conductivity tensor and $\mathbf{E}$ is the electric field, were reformulated as a Laplace equation for the electric potential $\mathbf{E} = -\nabla V$:
\begin{equation} \label{eq-LaplaceV}
    \nabla \cdot \left(\hat{\sigma} \nabla V \right) = 0.
\end{equation}
Equation \ref{eq-LaplaceV} was solved for the device geometry in Fig. \ref{reversed modulation} using a finite element method. A voltage drop was set across two edges of the device, and the boundary condition $\mathbf{J}\cdot \mathbf{\hat{n}} = 0$ was applied to the rest of the edges, where $\mathbf{\hat{n}}$ is the normal to the edge.

Fig. \ref{modulationSimulation} compares the expected current flow through an X-Y domain pattern in two cases: one where the conductivity is anisotropic and $\sigma_X$ and $\sigma_Y$ are related by a rotation, and one where the conductivity in each domain is isotropic, but its value changes between domains ($\hat{\sigma}^{X} = a\bm{1}_{2\times2}$ and   $\hat{\sigma}^{Y} = b\bm{1}_{2\times2}$), for example due to changes to the carrier density or dielectric constant across different domains. In the latter case, the current density modulations were independent of the direction of the flow: $J_{[11\bar{2}]}$ was lower on the X domains regardless of the direction of the device.
\subsection{Electrical transport measurements}

The nonlinear Hall effect was measured on a Hall cross device fabricated on the same sample as used for the SQUID imaging. The sample was anchored to a chip carrier by electrically- and thermally-conductive Ag-based epoxy, and the device was electrically contacted with Al wire bonds. Four-probe measurements were performed in a liquid He$^4$ cryostat at variable temperatures and magnetic fields (applied in the sample plane parallel to the electric current direction). Commercial lock-in amplifiers were used to apply an alternate current $I = 50$ \textmu A at a frequency $f = \omega/2\pi = 17.7$ Hz and detect the transverse first and second harmonic voltages while sweeping the magnetic field. A DC back-gate voltage was applied to tune the electronic density and mobility. The measured nonlinear Hall voltage was symmetrized and antisymmetrized with respect to the magnetic field $B$ to isolate distinct effects \cite{lesne2022designing}. The nonlinear conductivity shown in Fig. \ref{T-dep}c and Fig. \ref{Magnetotransport} corresponds to the $B$-symmetric component.
\subsection{Effective low-energy model}
An effective model for the (111) LaAlO$_3$/SrTiO$_3$ interface was derived using the theory of invariants. We use the $\ket{yz}, \ket{xz}, \ket{xy}$ basis for the Ti $t_{2g}$ orbitals. In the $\mathcal{C}_{3v}$ symmetry group, there is a $2\pi/3$ rotational symmetry and one mirror plane. 

Using the Gell-Mann matrices, we write the following Hamiltonian, which includes all terms, up to quadratic order, which respect the $\mathcal{C}_{3v}$ symmetry and preserve time reversal.
\begin{equation}
\begin{aligned}
    H & = \epsilon_{\text{trig}}(\Lambda_3+\Lambda_8/\sqrt{3}) + \\
    &(k_x^2+k_y^2)\left[c_1\Lambda_0 + c_2(\Lambda_3+\Lambda_8/\sqrt{3})\right] + \\
    &c_4\left[(k_x^2-k_y^2)\Lambda_1 + 2k_xk_y\Lambda_4\right] +\\
    &c_5\left[(k_x^2-k_y^2)(\Lambda_3/2 - \sqrt{3}/2\Lambda_8) + 2k_xk_y\Lambda_6\right]+\\
    &\alpha_{\text{OR}}(k_x\Lambda_5 + k_y\Lambda_2).
\end{aligned}
\end{equation}
The first term is the trigonal field splitting between the $a_{1g}$ and $e_g^{\pi}$ bands at the $\Gamma$ point. The last term is the orbital Rashba coupling, and the remaining terms account for the orbital and direction-depenent effective masses and for the trigonal warping. The prefactors were determined using the ARPES data of Ref. \cite{Rodel_2014}. 

The tetragonal distortion reduces the symmetry group to $\mathcal{C}_s$ by removing the rotational symmetry. This enables multiple additional terms in the Hamiltonian: splitting of the $e_g^{\pi}$ orbitals at the $\Gamma$ point, anisotropic linear terms (i.e. $k_x\Lambda_7$) and anisotropic quadratic terms (independent coefficients for the $k_x^2$ and $k_y^2$ terms). ARPES measurements do not resolve any of these contributions. To account for the anisotropy we only include a small (1 meV) splitting of the $e_g^\pi$ bands.

The Hamiltonian was diagonalized analytically and the conductivity was then computed according to
\begin{equation}
    \sigma_{\mu\nu} \sim \sum_{n=1}^{3} \int d^2k \frac{\partial \epsilon_n}{\partial k_\mu} \frac{\partial \epsilon_n}{\partial k_\nu} \left(- \frac{\partial f}{\partial E} \right)_{\epsilon = \epsilon_n (\mathbf{k})},
\end{equation}
where $f(\epsilon)$ is the Fermi-Dirac function and we assume the relaxation time is energy independent.

\section{Supplementary Materials}
This PDF file includes: Figs. S1 to S12.

\bibliography{References}
\bibliographystyle{unsrt}
\hfill
\newline
\textbf{Acknowledgments:} E.P., X.W. and B.K. were supported by the European Research Council Grant No. ERC-2019-COG-866236, the Israeli Science Foundation, grant no. ISF-228/22, COST Action CA21144, and the Pazy Research Foundation grant no. 107-2018.
G.S. acknowledges support from the Swiss National Science Foundation (grant no. PZ00P2\_223542). A.L. was supported by a Marie Sk{\l}odowska-Curie Individual Fellowship under grant MagTopCSL (ID 101029345) and by the Foundation for Polish Science through the IRA Programme co-financed by EU within SG OP.
C.O. acknowledges support from a VIDI grant (Project 680-47-543) financed by the Netherlands Organization for Scientific Research (NWO). E.L. acknowledges funding from the EU  Horizon 2020 research and innovation programme under the Marie Sk\l{}odowska-Curie grant agreement no. 707404.
\textbf{Author contributions:}
E.P., X.W. and B.K. designed and conducted the scanning SQUID experiments and finite element simulations. G.S. conducted the non-linear transport measurements. T.C.v.T. and E.L. fabricated and patterned the sample. A.L. and C.O. proposed the theoretical model. E.P, A.L. and C.O. performed the band structure calculations. E.P., X.W., C.O., A.D.C and B.K. discussed and interpreted the results with contributions from all co-authors. E.P., C.O., A.D.C and B.K. wrote the manuscript with input from all coauthors.
\textbf{Competing interests:}
The authors declare that they have no competing interests. 
\textbf{Data availability:} The raw data are available from the corresponding authors upon reasonable request.

\newpage

\onecolumngrid
\setcounter{section}{0}
\setcounter{figure}{0}
\renewcommand{\thefigure}{S\arabic{figure}}
\renewcommand{\theequation}{S.\arabic{equation}}
\renewcommand{\thetable}{S\arabic{table}}
\renewcommand{\thesection}{S\arabic{section}}

\renewcommand{\thefootnote}{\fnsymbol{footnote}}
\newpage
\begin{center}

\large{\textbf{SUPPLEMENTAL MATERIAL FOR\\
Imaging orbital Rashba induced charge transport anisotropy}}
\normalsize
\bigskip
\newline
Eylon Persky,$^{1,2}$ Xi Wang,$^{1,2}$ Giacomo Sala,$^{3}$ Thierry C. van Thiel,$^{4}$ Edouard Lesne,$^{4,5}$ Alexander
Lau,$^{6}$ Mario Cuoco,$^{7}$ Marc Gabay,$^{8}$ Carmine Ortix,$^{9*}$ Andrea D. Caviglia,$^{3\dagger}$ and Beena Kalisky$^{1,2\ddagger}$ 

$^1${\it Department of Physics, Bar-Ilan University, Ramat-Gan 5290002, Israel}\\
$^2${\it Institute of Nanotechnology and Advanced Materials, Bar-Ilan University, Ramat-Gan 5290002, Israel}\\
$^3${\it Department of Quantum Matter Physics, University of Geneva,
24 Quai Ernest Ansermet, CH-1211 Geneva, Switzerland}\\
$^4${\it Kavli Institute of Nanoscience, Delft University of Technology, Lorentzweg 1, 2628CJ Delft, Netherlands}\\
$^5${\it Max Planck Institute for Chemical Physics of Solids, 01187 Dresden, Germany}\\
$^6${\it International Research Centre MagTop, Institute of Physics, Polish Academy of Sciences, Aleja Lotnik\'{o}w 32/46, PL-02668 Warsaw, Poland}\\
$^7${\it Consiglio Nazionale delle Ricerche, CNR-SPIN, Italy}\\
$^8${\it Laboratoire de Physique des Solides, Universite Paris-Saclay, CNRS UMR 8502, Orsay, France}\\
$^9${\it Dipartimento di Fisica "E. R. Caianiello", Universit\'{a} di Salerno, IT-84084 Fisciano, Italy}\\
$^*${cortix@unisa.it}, $^\dagger${andrea.caviglia@unige.ch}, $^\ddagger${beena@biu.ac.il}

\bigskip

\begin{figure*}[h]
\centering
\includegraphics[scale=1]{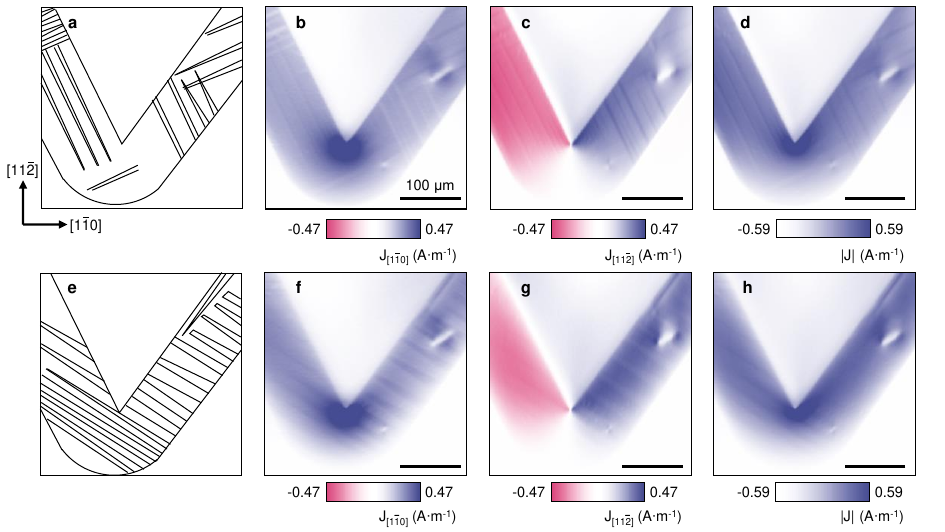}
\caption{Different domain patterns by thermal cycling. (a-d) Current flowing in a v-shape device with Y-Z domain patterns. (a) Sketch of domain patterns. (b-d) Inverted current density for $J_{[1\bar{1}0]}$, $J_{[11\bar{2}]}$, and $|J|$. (e-h) Current flowing in the same device but with X-Z domain patterns. Between these two sets of images, the temperature was cycled to 300 K. }
\label{T cycle}
\end{figure*}

\begin{figure*}[h]
\centering
\includegraphics[scale=1]{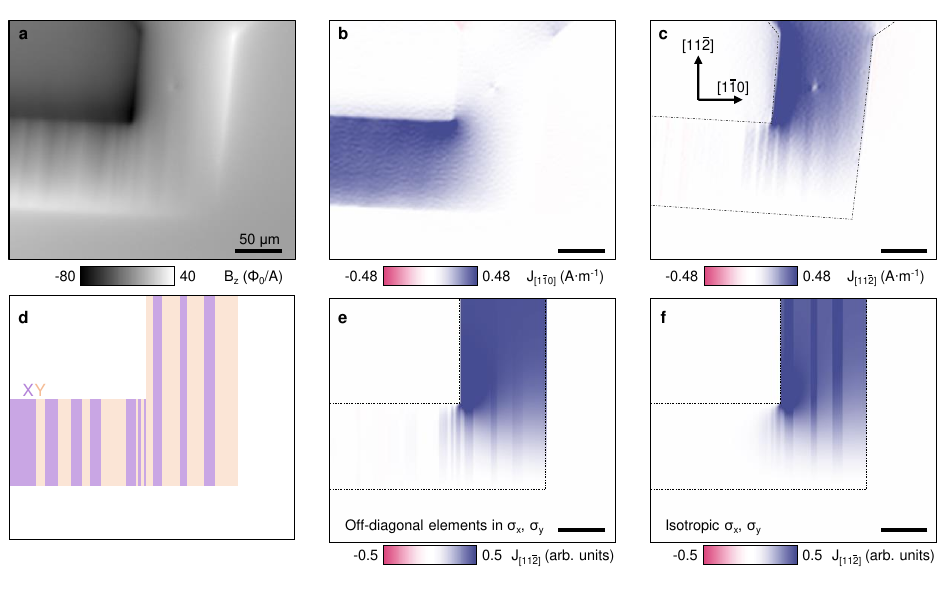}
\caption{Additional examples of changes in the modulations by reflecting the direction of current. (a) Raw magnetic flux data from an applied current of $40 \mathrm{\,\mu A}$ RMS. (b,c) The reconstructed $[1\bar{1}0]$ (b) and $[11\bar{2}]$ (c) components of the current density. (d) The expected XY domain configuration of this device. (e,f) The finite element simulations by using anisotropic conductivity (e) and isotropic conductivity (f). In panel f, the conductivity of X domains is set to be 20\% higher than Y domains.}
\label{example 1}
\end{figure*}

\begin{figure*}[h]
\centering
\includegraphics[scale=1]{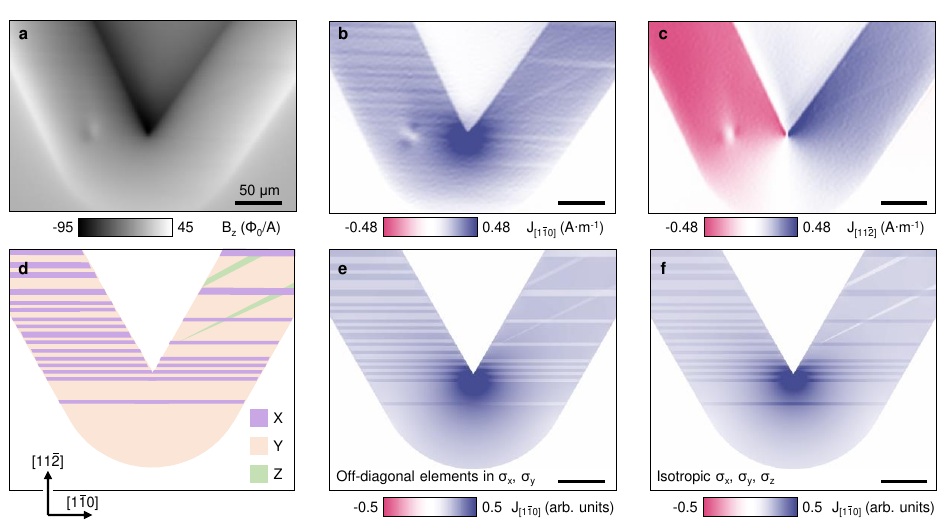}
\caption{Additional examples of changes in the modulations by reflecting the direction of current. This is the same device shown in Fig. \ref{reversed modulation} but the patterns are different due to a thermal cycle to 300 K. (a) Raw magnetic flux data from an applied current of $40 \mathrm{\,\mu A}$ RMS. (b,c) The reconstructed $[1\bar{1}0]$ (b) and $[11\bar{2}]$ (c) components of the current density. (d) The expected XYZ domain configuration of this device. (e,f) The finite element simulations by using anisotropic conductivity (e) and isotropic conductivity (f). In panel f, the conductivity of X and Z domains is set to be 20\% higher or weaker than Y domains.}
\label{example 2}
\end{figure*}

\begin{figure*}[h]
\centering
\includegraphics[scale=1]{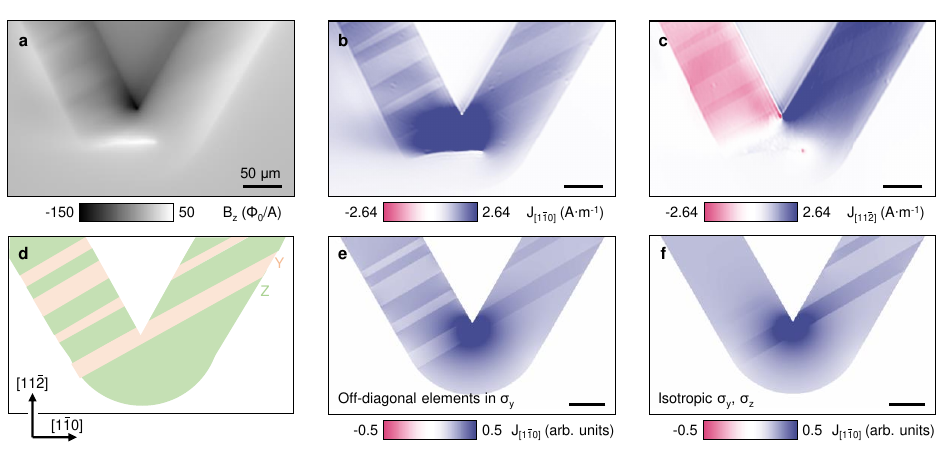}
\caption{Additional examples of changes in the modulations by reflecting the direction of current. (a) Raw magnetic flux data from an applied current of $264 \mathrm{\,\mu A}$ RMS. (b,c) The reconstructed $[1\bar{1}0]$ (b) and $[11\bar{2}]$ (c) components of the current density. (d) The expected YZ domain configuration of this device. (e,f) The finite element simulations by using anisotropic conductivity (e) and isotropic conductivity (f). In panel f, the conductivity of Y domains is set to be 20\% higher than Z domains.}
\label{example 3}
\end{figure*}

\begin{figure*}[h]
\centering
\includegraphics[scale=1]{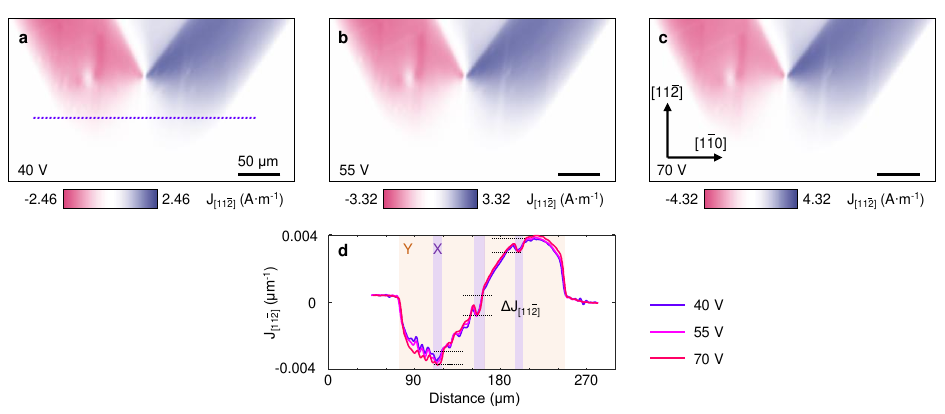}
\caption{Gate dependence of $J_{[11\bar{2}]}$ in the device shown in Fig. \ref{reversed modulation} at 7 K. (a-c) From the gate voltage of 40 to 70 V, the relevant currents are 205, 277, 360 $\mu A$, correspondingly. The modulation amplitudes of these three images are plotted in Fig. \ref{band structure}f, light orange color. We used a constant voltage bias setup with a lock-in amplifier, accounting for this change in current with the gate voltage. The data is normalized by the applied current. (d) Line cuts taken from panels a, b, and c. The data is normalized by the applied current.}
\label{Gate dep images}
\end{figure*}

\begin{figure*}[h]
\centering
\includegraphics[scale=1]{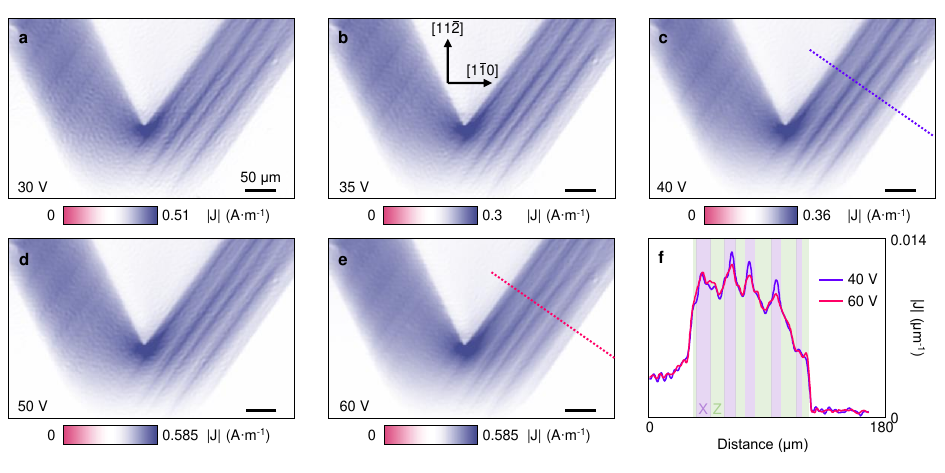}
\caption{Gate dependence of $|J|$ at 5.5 K, in the device shown in the Extended Data Fig. \ref{current inversion 2} and  \ref{modulation}. (a-e) From the gate voltage of 30 to 60 V, the relevant currents are 34, 20, 24, 39, 39 $\mu A$, correspondingly. The modulation amplitudes of these XZ domain structures are plotted in Fig. \ref{band structure}f, dark blue color. (f) Line cuts taken from panels c and e. The modulation at 40 V is slightly stronger than at 60 V. The data is normalized by the applied current.}
\label{Gate dep images 2}
\end{figure*}

\begin{figure*}[h]
\centering
\includegraphics[scale=1]{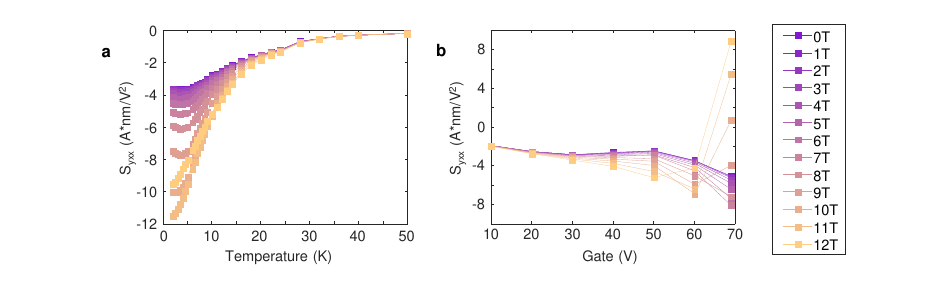}
\caption{Magnetotransport data. (a) Temperature and (b) gate dependence of the non-linear conductivity up to 12 T, showing similar onset temperature to the anisotropy, and a non-monotonic dependence on the gate voltage. The data at zero field is presented in the main text Fig. \ref{T-dep} panels c and e.}
\label{Magnetotransport}
\end{figure*}

\begin{figure*}[h]
\centering
\includegraphics[scale=1]{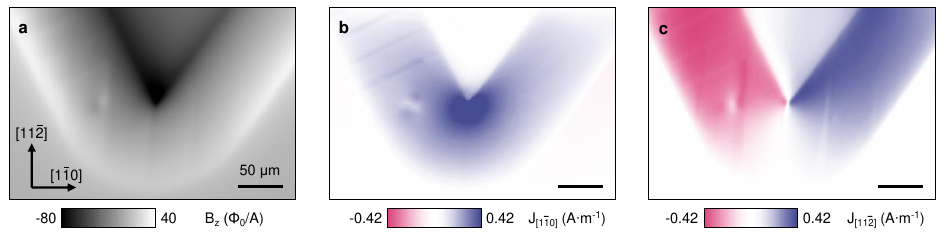}
\caption{Current density reconstruction of the device shown in Fig. \ref{reversed modulation} at gate voltage 70 V. (a) Raw magnetic flux data due to an applied current of $34 \mathrm{\,\mu A}$ RMS in a $100 \mathrm{\,\mu m}$ wide (111)-LAO/STO device. (b,c) The reconstructed $[1\bar{1}0]$ (b) and $[11\bar{2}]$ (c) components of the current density.}
\label{current inversion 1}
\end{figure*}

\begin{figure*}[h]
\centering
\includegraphics[scale=1]{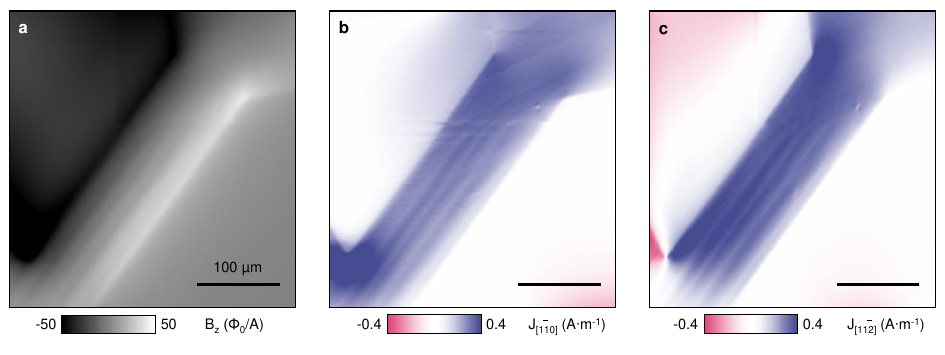}
\caption{Current density reconstruction. (a) Raw magnetic flux data due to an applied current of $40 \mathrm{\,\mu A}$ RMS in a $100 \mathrm{\,\mu m}$ wide (111)-LAO/STO device. (b,c) The reconstructed $[1\bar{1}0]$ (b) and $[11\bar{2}]$ (c) components of the current density. All data are normalized by the applied current.}
\label{current inversion 2}
\end{figure*}

\begin{figure*}[p]
\centering
\includegraphics[scale=1]{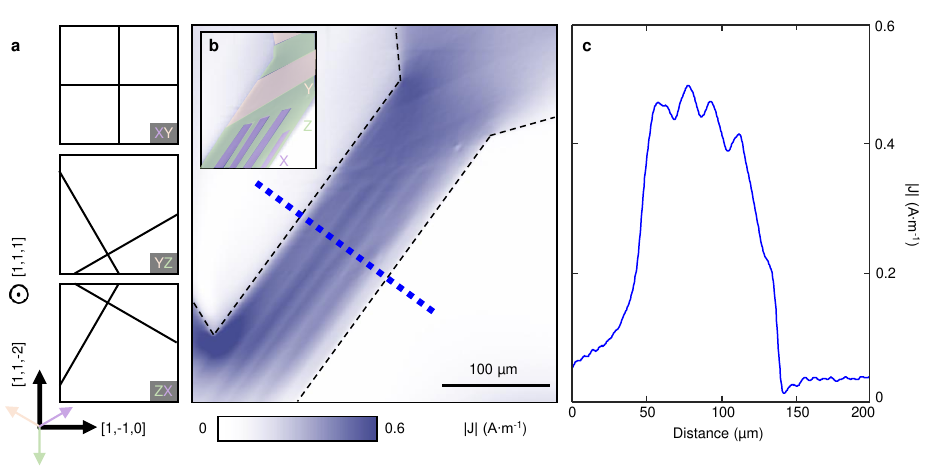}
\caption{Tetragonal domain patterns in $(111)$ LAO/STO. (a) domain walls projected onto the $(111)$ oriented surface. Boundaries between X and Y domains form $0^{\circ}$ and $90^{\circ}$ with respect to the $[1\bar{1}0]$ direction, Y-Z boundaries form $30^{\circ}$ and $120^{\circ}$ and X-Z boundaries form $60^{\circ}$ and $150^{\circ}$. (b) A representative current density image of a patterned $(111)$ LAO/STO device, showing modulations along wide stripes oriented $30^{\circ}$ and $60^{\circ}$. Inset: From these orientations, we infer the structural domain configuration in this area. The black dashed lines mark the edges of the device. (c) Line cut of the current density taken along the blue dashed line in b, showing the X-Z domain pattern causes a current density modulation of 11\%.}
\label{modulation}
\end{figure*}

\makeatletter
\@fpsep\textheight
\makeatother

\begin{figure*}[h]
\centering
\includegraphics[scale=1]{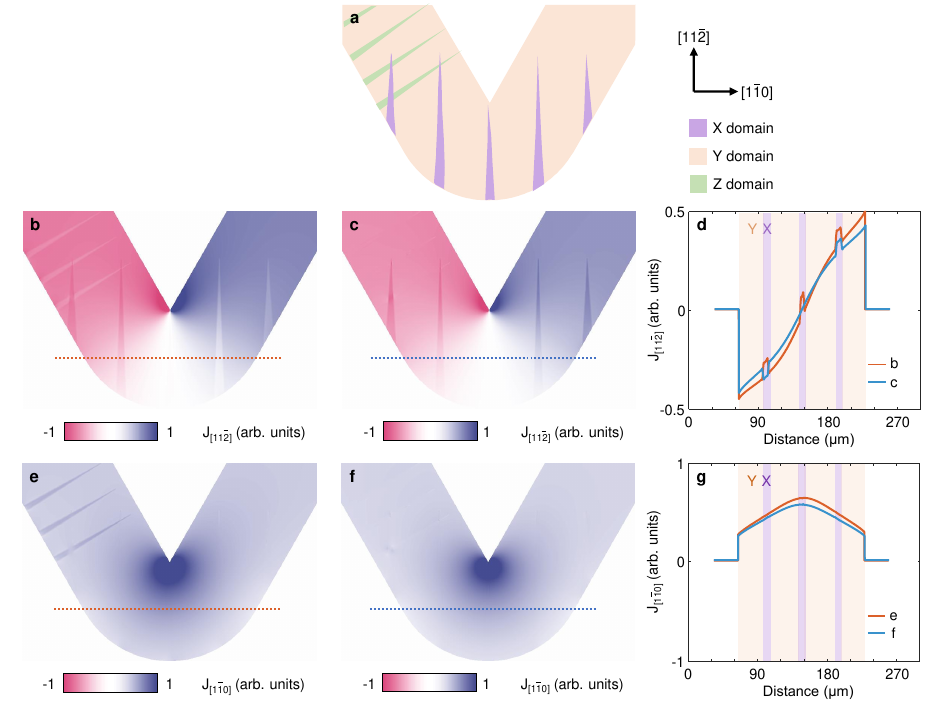}
\caption{Finite element simulations of current flow in a triangle shape device with an X-Y domain pattern. (a) The domain patterns. (b) The modulations in $J_{[11\bar{2}]}$ are caused by anisotropic conductivity. (c) The modulations are caused by different isotropic conductivities in the different domains. Note that the latter scenario does not reproduce the reversed modulations observed in the experiment or the modulation in Z domains. (d) Line cuts from the current density maps in panels b, c. (e) Simulation results with anisotropic conductivity in $J_{[1\bar{1}0]}$ component of panels a, b where the X-Y domain modulation is barely visible. (f) In the simulation of different isotropic conductivities, the modulation in Z domains is not visible as well. (g) Line cuts from the current density maps in panels e, f.}
\label{modulationSimulation}
\end{figure*}

\begin{figure*}[h]
\centering
\includegraphics[scale=1]{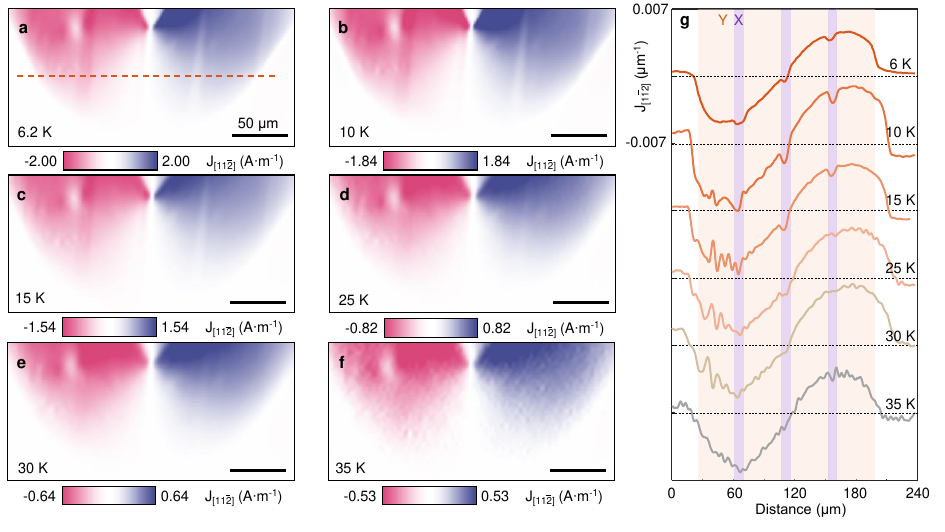}
\caption{Temperature dependence of $J_{[11\bar{2}]}$ in the device shown in Fig. \ref{reversed modulation} at gate voltage 70 V. (a-f) From 6 to 35 K, the amplitudes of current are 167, 153, 128, 68, 53, 44 $\mu A$, correspondingly. We used a constant voltage bias setup with a lock-in amplifier, accounting for the change of current with temperatures. (g) Line cuts taken from panels a-f. At 35 K the modulation is beyond our sensitivity. The data is normalized by the applied current. }
\label{T dep images}
\end{figure*}

\clearpage
\end{center}
\end{document}